\documentstyle[12pt,epsf]{article}
\newcommand{\beq}{\begin{equation}}
\newcommand{\eeq}{\end{equation}}
\newcommand{\bea}{\begin{eqnarray}}
\newcommand{\eea}{\end{eqnarray}}
\newcommand{\un}{\underline}
\newcommand{\half}{{\scriptstyle{{1\over 2}}}}

\newcommand{\real}{\relax{\rm I\kern-.18em R}}

\newcommand{\id}{\mbox{$id$}}

\newcommand{\sgbar}{\bar{\sg}}
\newcommand{\tr}{\mbox{\,tr\,}}
\newcommand{\ch}{{\hat{c}}}
\newcommand{\dh}{{\hat{d}}}
\newcommand{\wdrie}{\sqrt{3}}
\newcommand{\al}{\alpha}

\newcommand{\gm}{\gamma}

\newcommand{\dl}{\delta}
\newcommand{\Dl}{\Delta}

\newcommand{\fie}{\varphi}

\newcommand{\sg}{\sigma}

\newcommand{\Ss}[1]{\mbox{$\cal #1$}}
\newcommand{\pr}{\partial}

\newcommand{\Order}[1]{\Ss{O}\left(#1\right)}

\newcommand{\fm}{\mbox{fm}}

\begin{document}
\vskip-1cm
\hfill INLO-PUB-10/95
\vskip5mm
\begin{center}
{\LARGE{\bf{\underline{Glueball spectroscopy on $S^3$}}}}\\
\vspace*{5mm}
\vspace*{1cm}{\large Bas van den Heuvel\footnote{e-mail:
bas@rulgm0.LeidenUniv.nl}} \\
\vspace*{1cm}
Instituut-Lorentz for Theoretical Physics,\\
University of Leiden, PO Box 9506,\\
NL-2300 RA Leiden, The Netherlands.\\
\end{center}
\vspace*{5mm}{\narrower\narrower{\noindent
\underline{Abstract:}  For SU(2) gauge theory on the three-sphere
we implement the influence of the boundary of the fundamental domain,
and in particular the $\theta$-dependence,
on a subspace of low-energy modes of the gauge field.
We construct a basis of functions that respect these boundary conditions
and use these in a variational approximation of the spectrum of
the lowest order effective hamiltonian.
}\par}

\section{Introduction}
It is our aim to study the dynamics of the low-energy modes
of pure SU(2) gauge theory defined in a finite volume
\cite{baa2,baa1,baa6}. If we take the volume small, the mechanism of
asymptotic freedom results in a small coupling constant and we can
use standard perturbation theory.
Gradually increasing the volume then allows us to monitor the onset of
non-perturbative phenomena.
Especially when the non-perturbative effects manifest themselves
appreciably only in a small number of low-lying energy modes, this can be
described adequately using a hamiltonian formulation.
We are interested in the influence of the multiple vacuum structure of
the theory on the glueball spectrum: in particular we would like
to see the dependence of the energies on the $\theta$-angle which
corresponds to the transition over instanton barriers.
\par
In our approach, we impose the Coulomb gauge by restricting the
gauge fields to a so-called fundamental domain~\cite{sem,zwan}.
This is a convex subset of the space of all gauge field configurations
that is in one-to-one correspondence (modulo constant gauge transformations)
with the space of gauge orbits. The latter
is precisely the physical configuration space on which we want to study
the dynamics.
\par
Our strategy will be to split up the gauge field in orthogonal modes
and to reduce the dynamics of this infinite number of degrees of freedom
to a quantum mechanical problem with a finite number of modes.
We do this by taking standard harmonic oscillator wave functions for the
high-energy modes of the field.
For the low-energy modes, the onset
of non-perturbative effects means that the wave functional
starts to spread out over configuration space and will, in particular,
become sensitive to conditions imposed on the boundary of the
fundamental domain. We therefore replace the wave functionals for these
modes by functions that satisfy the ($\theta$-dependent) boundary conditions.
\par
The dynamics of these modes can be regarded as
an effective low-energy theory. Similar
to a Born-Oppenheimer approximation we integrate out the fast (high-energy)
modes and are left with an effective theory of the slow modes. If
we are at energies at which the higher modes cannot be excited,
these modes will, through virtual processes, in first order only result in a
renormalisation of the coupling  constant.
\par
To make contact with lattice calculations it would be most natural
to take the finite (spatial) volume to be a 3-torus $T^3$.
Detailed knowledge of the vacuum structure in this case is however limited.
In particular the instantons, which are
the gauge field configurations that describe tunnelling between
different vacua, are only known numerically~\cite{gar}.
To circumvent this problem
we take our space to be the three-sphere $S^3$
on which the instantons are known analytically.
\par
Previous studies of gauge theory on $S^3$~\cite{cut1} did not
use conditions at the boundary of the fundamental domain,
but focussed on the influence of the Gribov horizon:
rescaling the wave function with the square
root of the Faddeev-Popov determinant results in a potential
with an infinite barrier at the Gribov horizon.
\par
In this paper, we first construct suitable boundary conditions
on the space of low-energy modes and then give a basis of functions
that respect these boundary conditions. Using this basis we
perform a Rayleigh-Ritz analysis to approximate the spectrum of
the lowest order effective hamiltonian.
The obtained energy differences give a first hint at the values
of the glueball masses in the different (scalar, tensor) sectors.
\section{The Effective Theory}
We will briefly review the formalism developed in~\cite{baa1,baa6}
for gauge fields on $S^3$ and construct the effective hamiltonian.
We embed $S^3$ in $\real^4$ by considering the unit sphere parametrized by a
unit vector $n_\mu$. Using the scale-invariance of the classical
hamiltonian, we can make the restriction to a sphere of radius $R=1$.
$R$-dependence can be reinstated on dimensional grounds.
We introduce the unit quaternions $\sg_\mu$ and their
conjugates $\bar{\sg}_\mu = \sg^\dagger_\mu$ by
\beq
  \sg_\mu = ( \id ,i \vec{\tau}), \hspace{1.5cm}
  \bar{\sg}_\mu = ( \id,- i \vec{\tau}).
\eeq
They satisfy the multiplication rules
\beq
 \sg_\mu \sgbar_\nu = \eta^\al_{\mu \nu} \sg_\al, \hspace{1.5cm}
  \sgbar_\mu \sg_\nu = \bar{\eta}^\al_{\mu \nu} \sg_\al,
\eeq
where we used the 't Hooft $\eta$ symbols~\cite{tho}, generalised slightly to
include a component symmetric in $\mu$ and $\nu$ for $\al=0$.
We can use $\eta$ and $\bar{\eta}$ to define orthonormal framings of $S^3$,
which were motivated by the particularly simple form of the instanton
vector potentials in these framings. The framing for $S^3$ is
obtained from the framing of $\real^4$ by restricting in the following
equation the four-index $\al$ to a three-index $a$
(for $\al = 0$ one obtains the normal on $S^3$):
\beq
  e^\al_\mu = \eta^\al_{\mu \nu} n_\nu , \hspace{1.5cm}
  \bar{e}^\al_\mu = \bar{\eta}^\al_{\mu \nu} n_\nu.
\eeq
A gauge field on $S^3$ ($n_\mu A_\mu = 0$)
can be written with respect to either framing, e.g.
\beq
   A_\mu= A_i e_\mu^i = A^a_i e_\mu^i \frac{\sg_a}{2}
\eeq
A gauge transformation $g$ acts as follows:
\beq
  \left(^g A\right)_i = g^{-1} A_i g + g^{-1} \pr_i g = g^{-1} D^{(A)}_i g.
\eeq
In order to isolate the lowest energy levels, we define the quadratic
fluctuation operator \Ss{M} by
\beq
  \Ss{V}(A) =  - \int_{S^3} \frac{1}{2} \tr(F_{ij}^2)
  = \int_{S^3} \tr(A_i \Ss{M}_{ij} A_j) + \Order{A^3}. \label{fluctdef}
\eeq
The $18$ dimensional space $A(c,d)$ given by
\beq
    A_\mu(c,d) = \left(c^a_i  e^i_\mu+d^a_j\bar{e}^j_\mu \right)
  \frac{\sg_a}{2} \label{Acddef}
\eeq
is the eigenspace of \Ss{M} corresponding to its lowest eigenvalue 4,
whereas the next eigenvalue is 9.
As was explained in~\cite{baa1,baa6}, this
space contains gauge copies of the vacuum $A=0$ and the tunnelling paths
inbetween these vacua. In particular, the vacuum $A_a = - \sg_a$
($c^a_i = -2 \dl^a_i,~d^a_i = 0$) is a copy of $A=0$
under the gauge transformation
$g = n \cdot \sgbar$ with winding number one. The tunnelling path
is $c^a_i = -u \dl^a_i, d^a_i = 0$ with $u$ running from $0$ to $2$.
For $u=1$ it passes through the sphaleron, which is a saddle point of the
energy functional. The vacuum $c^a_i = 0,~d^a_i = -2 \dl^a_i$ is a
copy of $A=0$ under $g = n \cdot \sg$. The corresponding saddle point is
called "anti-sphaleron" and is actually a copy of the
sphaleron under this gauge transformation.
The energy functional for these 18 modes is given by
\beq
  \Ss{V}(c,d) \equiv - \int_{S^3} \frac{1}{2} \tr(F_{ij}^2)
  = \Ss{V}(c) + \Ss{V}(d) + \frac{2 \pi^2}{3}
   \left\{ (c^a_i)^2 (d^b_j)^2 - (c^a_i d^a_j)^2 \right\}\label{pot}
,\eeq
\beq
  \Ss{V}(c) = 2 \pi^2 \left\{ 2 (c^a_i)^2 + 6 \det c +
  \frac{1}{4}[(c^a_i c^a_i)^2 - (c^a_i c^a_j)^2 ] \right\}.
\eeq
The lowest order hamiltonian for these modes is
\beq
  R~\Ss{H}(c,d) = -\frac{f}{2} \left(
  \frac{\pr^2}{\pr c^a_i \pr c^a_i} + \frac{\pr^2}{\pr d^a_i \pr d^a_i}
   \right) + \frac{1}{f} \frac{\Ss{V}(c,d)}{2 \pi^2} \label{hamdef}
\eeq
with $f = \frac{g(R)^2}{2 \pi^2}$ and $g(R)$ the renormalized
coupling constant.
\par
The boundary of the fundamental domain in the full $(c,d)$ space
is not known, but at the energies we are interested in, the boundary
conditions are relevant only at those points where the potential is low,
i.e.\ at the sphalerons. At other
boundary points the potential will be (much)
higher then the energy of the wave functional. This means that the
wave functional will have decayed exponentially at these points
and that the precise boundary conditions
should not have a big influence on the spectrum. By the same token,
the precise location of the boundary in these regions is not
important either. This gives us the freedom to choose tractable
boundary conditions.
\par
At the sphalerons however, the boundary conditions are fixed.
Since the gauge transformation connecting the two sphalerons has
winding number one, we have to set
\beq
  \Psi(A(\mbox{Sph},0)) = e^{i \theta} \Psi(A(0,\mbox{Sph})),
\label{oerbc}
\eeq
thus introducing the $\theta$-angle.

\section{The Rayleigh-Ritz basis}
We now want to determine the spectrum of the hamiltonian~(\ref{hamdef}),
properly restricted to the fundamental domain. The variational basis we need
must incorporate the boundary conditions but also respect
the symmetries of the hamiltonian to obtain an optimal block diagonalisation.
\par
We first turn to the boundary conditions.
We define radial coordinates $r_c$ and $r_d$
by
\beq
  r_c = \left[ c^a_i c^a_i \right]^\half, \quad
  r_d = \left[ d^a_i d^a_i \right]^\half.
\eeq
The sphaleron has radial coordinates $(\sqrt 3,0)$ and angular coordinates
$\ch^a_i = - \dl^a_i$ (with $\ch^a_i = c^a_i/r_c$). It will be connected
with the anti-sphaleron at $(0,\sqrt 3)$.
We will restrict the $(r_c,r_d)$ plane by $r_c < \wdrie,~r_d < \wdrie$
and impose boundary conditions at the edges.
This means that we will be working towards basis functions of the
form $\phi(r_c,r_d) Y(\ch,\dh)$.
\par
As was explained earlier, for values of the coupling constant at
which our approximation will
be valid, only the effect of the boundary conditions at the sphaleron
will be felt. By imposing boundary conditions in the $(r_c,r_d)$ plane
we pair up two submanifolds, of which only the sphaleron/anti-sphaleron
need belong to the boundary of the fundamental domain.
\par
Consider the following decomposition of the full wave function
\beq
  \Psi = \frac{1}{r_c^4 r_d^4} \sum_n \psi^{(n)}(c,d) \chi_{[c,d]}^{(n)}(q),
\eeq
where $q$ denotes all the modes orthogonal to the $c$ and $d$ modes.
The functions $\chi_{[c,d]}^{(n)}(q)$ are chosen to be
(perturbative) eigenstates of the transverse hamiltonian.
In the adiabatic approximation we assume the transverse wave function
to be in its ground state and we assume this ground state
to decouple dynamically from the excited states.
This results in a hamiltonian for $\psi = \psi^{(0)}$
given by
\beq
  \Ss{H} = -\frac{f}{2} \left(
  \frac{\pr^2}{\pr r_c^2} + \frac{\pr^2}{\pr r_d^2} -
  12 \left(\frac{1}{r_c^2} + \frac{1}{r_d^2}\right) +
  \frac{1}{r_c^2} \Dl_\ch + \frac{1}{r_d^2} \Dl_\dh
   \right) +
   \frac{1}{f} \frac{\Ss{V}(c,d)}{2 \pi^2} \label{hamrad}
\eeq
with $\Dl_\ch$ the laplacian in the angular coordinates.
The boundary condition on $\psi$ follows directly from~(\ref{oerbc}):
\beq
  \psi(\mbox{Sph},0) = e^{i \theta} \psi(0,\mbox{Sph}).
  \label{bcgen1}
\eeq
To obtain the condition on the derivative of $\psi$ we must take
the residual constant gauge symmetry into account.
Focussing on the sphaleron path $c^a_i = -u \dl^a_i$, we take coordinates
$c = S(\vec{\al}) H(u,h_i)$, with $S  \in \mbox{SO}(3)$ and $H$ symmetric.
Under a constant gauge transformation only $S$ changes:
$H(u,h_i)$ forms a set of gauge invariant curvilinear
coordinates. Matching along the sphaleron path across the boundary
of the fundamental domain, we need to compensate for the curvature
with the appropriate jacobian factor $u^{3/2}$ (details will
appear elsewhere),
which results in the following boundary condition on $\psi$:
\beq
  \frac{\pr (r_c^{-5/2}\psi)}{\pr r_c}(\mbox{Sph},0) =
    - e^{i \theta} \frac{\pr (r_d^{-5/2}\psi)}{\pr r_d}(0,\mbox{Sph}).
\label{bcgen2}
\eeq
\begin{table}[t]
\begin{center}
\begin{tabular}{||l|l||l|l||} \hline
\multicolumn{2}{||c||}{$\langle \ch | L;l_s,l_r,\tau;m_s,m_r \rangle$} &
\multicolumn{2}{|c||}{$\langle \ch \dh |
  j,m,l_s;L_1,l_1,\tau_1;L_2,l_2,\tau_2 \rangle$} \\ \hline
Operator & Eigenvalue & Operator & Eigenvalue \\ \hline
$(\vec{L}^S_c)^2$ & $l_s ( l_s + 1) $ & $(\vec{L}^S_c)^2$ & $l_s ( l_s + 1) $
\\
$(\vec{L}^S_c)_3$ & $m_s $ & $(\vec{L}^S_d)^2$ & $l_s ( l_s + 1) $ \\
&& $(\vec{J}^S)^2$ & $0$ \\
&& $\vec{J}^S_3$ & $0$ \\ \hline
$(\vec{L}^R_c)^2$ & $l_r ( l_r + 1) $ & $(\vec{L}^R_c)^2$ & $l_1 ( l_1 + 1) $
\\
$(\vec{L}^R_c)_3$ & $m_r $ & $(\vec{L}^R_d)^2$ & $l_2 ( l_2 + 1) $ \\
&& $(\vec{J}^R)^2$ & $j (j+1)$ \\
&& $\vec{J}^R_3$ & $m$ \\ \hline
$\Dl_\ch$ & $ - L ( L + 7) $ & $\Dl_\ch$ & $ - L_1 ( L_1 + 7) $ \\&& $\Dl_\dh$
& $ - L_2 ( L_2 + 7) $ \\ \hline
\end{tabular}
\end{center}
\caption{Behaviour of the functions of the angular variables}
\label{angtabel}
\end{table}
We now turn to the symmetries of the hamiltonian. $\Ss{H}(c,d)$ is invariant
under the transformation $ c \rightarrow S c R_1,~d \rightarrow S d R_2$
with $S,R_1,R_2 \in \mbox{SO}(3)$ and under the
interchange $c \leftrightarrow d$.
The generators of left- and right
multiplication are $\vec{L}^R_c$, $\vec{L}^S_c$, $\vec{L}^R_d$
and $\vec{L}^S_d$. The following set of operators commutes:
\beq
 \left\{\Ss{H},\vec{J}^S,\vec{J}^R,(\vec{L}^R_c)^2 + (\vec{L}^R_d)^2,
  (\vec{L}^S_c)^2 + (\vec{L}^S_d)^2,\Ss{P}\right\}.
\eeq
\Ss{P} is defined by $\Ss{P} f(c,d) \equiv f(d,c)$. On $S^3$ it
corresponds to the parity $(n_0,\vec{n}) \leftrightarrow (n_0,-\vec{n})$.
The operator $\vec{J}^S \equiv \vec{L}^S_c + \vec{L}^S_d$ implements
constant gauge transformations: we have to demand $(\vec{J}^S)^2 = 0$
for physical wave functions.
The operator $\vec{J}^R \equiv \vec{L}^R_c + \vec{L}^R_d$ is
the rotation operator. The different sectors under this symmetry
correspond to scalar glueballs, vector glueballs, etc.
\par
An orthonormal basis of functions of $\ch$ is given by the set
$\left\{\langle \ch|L;l_s,l_r,\tau;m_s,m_r \rangle \right\}$. Each of these
functions is a certain polynomial in $\ch$, homogeneous of degree $L$.
Its eigenvalues under the various symmetries are collected in
table~\ref{angtabel}.
A possible degeneracy is labeled by $\tau$.
Using algebraic manipulation programs, we were able to construct
these functions explicitely for $L \leq 10$.
Using Clebsch-Gordan coefficients,
we define a function $Y^{i_1i_2}(\ch,\dh)$ which is an eigenfunction of
$\vec{J}^R$ and of $\vec{J}^S$ ($i$ denotes a set $(L,l_s,l_r,\tau)$):
\bea
Y^{i_1i_2}(\ch,\dh) &=& \langle \ch \dh | j,m,i_1,i_2 \rangle  \nonumber \\
  &=& \sum_{m_s = -l_s}^{l_s} \sum_{m_1 = -l_1}^{l_1} \sum_{m_2 = -l_2}^{l_2}
  (-1)^{l_1 - l_2 + m} \sqrt{2 j + 1}
   \left( \begin{array}{ccc}
     l_1 & l_2 & j \\
     m_1 & m_2 & -m
   \end{array} \right)  \nonumber \\
  & &(-1)^{l_s - m_s} \frac{1}{\sqrt{2 l_s + 1}}
    \langle \ch | L_1;l_s,l_1,\tau_1;m_s,m_1 \rangle
       \langle \dh | L_2;l_s,l_2,\tau_2;-m_s,m_2 \rangle. \label{cddef}
\eea
Note that in order to have $\vec{J}^S = 0$, the functions of $c$ and $d$
need to have the same $l_s$. This restricts the possible combinations of
$i_1$ and $i_2$.
\par
Let us focus on the case $\theta = 0$. In this case we can in principle take
the basis functions to be the eigenfunctions of the kinetic part
of the hamiltonian. These functions are given by
\beq
  \fie^{(L_1)}_{\gm_1}(r_c)  \fie^{(L_2)}_{\gm_2}(r_d)
   Y^{i_1i_2}(\ch,\dh),
\eeq
where $L_k$ is determined by $i_k$ and
$\fie^{(L)}_\gm(r) = \gm r j_{(3+L)}(\gm r)$
with $j_p(z)$ the spherical Bessel function of order $p$.
During the variational stage of the calculation, however,
the use of Bessel functions of different order will lead to a
large number of integrals. Therefore we take the radial functions to
be independent of $L_1$ and $L_2$ and define
\beq
\psi^{i_1i_2}_{\gm_1\gm_2}(c,d) =
 \fie_{\gm_1}(r_c)  \fie_{\gm_2}(r_d) Y^{i_1i_2}(\ch,\dh),
\eeq
with $\fie_\gm(r) = \gm r j_3(\gm r)$. The functions
$\psi^{i_1i_2}_{\gm_1\gm_2}(c,d)$ now have singularities at
certain points, but these are not felt during the variational calculation.
Taking even and odd combinations gives
\beq
\psi^{pi_1i_2}_{\gm_1\gm_2}(c,d) =
 \psi^{i_1i_2}_{\gm_1\gm_2}(c,d) + p~\psi^{i_1i_2}_{\gm_1\gm_2}(d,c)
\eeq
We implement the boundary conditions~(\ref{bcgen1}) and (\ref{bcgen2})
for $\theta = 0$ by imposing
the following conditions on $\gm_1$ and $\gm_2$:
\bea
  p = -1 &:&  \fie_{\gm_1}(\wdrie) = \fie_{\gm_2}(\wdrie) = 0, \\
  p = 1 &:&  \frac{\pr ( r^{-5/2} \fie_{\gm_1})}{\pr r} (\wdrie) =
             \frac{\pr ( r^{-5/2} \fie_{\gm_2})}{\pr r} (\wdrie) = 0.
\label{bcspec}
\eea
These conditions are expected to be accurate as long as the
wave function transverse to the sphaleron path (near the sphalerons) is
predominantly in its ground state.
For general $\theta$ we multiply $\psi^{pi_1i_2}_{\gm_1\gm_2}(c,d)$
with a phase factor $\exp(i \theta \alpha(r_c,r_d))$. The function $\al$
is a kind of Cherns-Simons functional that gives the right behaviour to
the wave function under large gauge transformations.
The resulting
functions no longer have well-defined parity, but they do obey the
general boundary conditions for suitable $\alpha$.
Also the hermiticity of the hamiltonian for these
functions can be checked explicitely.
Sufficient conditions
on $\alpha$ are: $\alpha(r_c,r_d) = - \alpha(r_d,r_c)$ and
$\alpha(\wdrie,0) = \half$. We choose
\beq
   \alpha(r_c,r_d) = \frac{1}{2} \left(
     \left(\frac{r_c}{\wdrie} \right)^\beta -
     \left(\frac{r_d}{\wdrie} \right)^\beta \right). \label{alphadef}
\eeq
For $\beta \rightarrow \infty$ we approach the situation that
the phase factor over the entire edge is constant
and equal to $e^{i \theta}$. But already for the choice $\beta = 2$,
the boundary conditions at the sphalerons are taken into account properly.
\section{The Rayleigh-Ritz analysis}
To perform the Rayleigh-Ritz analysis, we take a finite subset
of the basis constructed above. We then calculate the matrix of \Ss{H} with
respect to this truncated basis and diagonalise.
The eigenvalues found are upper bounds for the true
eigenvalues. In order to get lower bounds, we use
Temple's inequality~\cite{ree} which as input requires the expectation values
of $\Ss{H}^2$ for the variational eigenvectors.
For the calculation of the matrix elements we essentially need the
radial integrals
\beq
  \int_0^{\wdrie} dr~r^n \fie_{\gm'}(r) \fie_{\gm}(r)
\eeq
with $\gm$ and $\gm'$ given by the boundary conditions~(\ref{bcspec}).
The angular sector is more difficult: especially $\Ss{V}^2$ contains
some non-trivial angular operators.
By virtue of the construction~(\ref{cddef}) all the angular integrations
over $\ch$ and $\dh$ can be reduced to integrations over just
$\ch$. Both the radial and the angular integrals were tabulated
and stored for later use in a Fortran program.
\par
Due to the symmetries, the function space splits up in different sectors
characterized by the conserved quantum numbers $j$, $m$,
$ l_1 (l_1+1 ) + l_2 (l_2 + 1)$ and, for $\theta = 0$, $p$.
It is an amusing fact that the only influence of the conserved
numbers $j$ and $m$ on the spectrum is through the number $(-1)^j$ and through
the fact that $l_1,l_2$ and $j$ have to satisfy a triangle inequality.
If we denote a sector with a triple $(l_1,l_2,j)$,
this for instance means that the sectors $(1,1,0)$ and $(1,1,2)$
are degenerate.
\par
\begin{figure}[t]
\epsfxsize=152mm
\epsffile{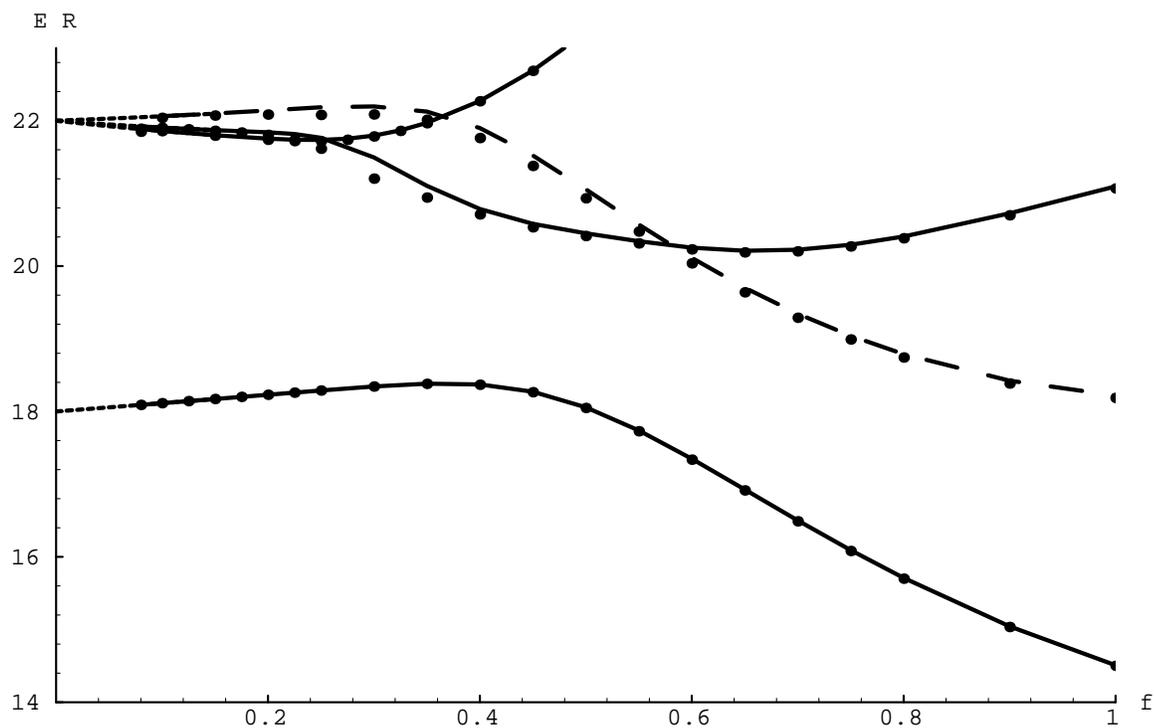}
\caption{Lowest energy levels for $\theta = 0$.
Drawn curves correspond to levels
in the $(0,0,0)$ sector. The dashed curve denotes the ground level in the
$(1,1,2)$-even sector. The short-dashed curves are the
perturbative expansions, and the individual dots are lower
bounds on the levels as obtained by Temple's inequality.}
\label{niveaux}
\end{figure}
\begin{figure}[t]
\epsfxsize=152mm
\epsffile{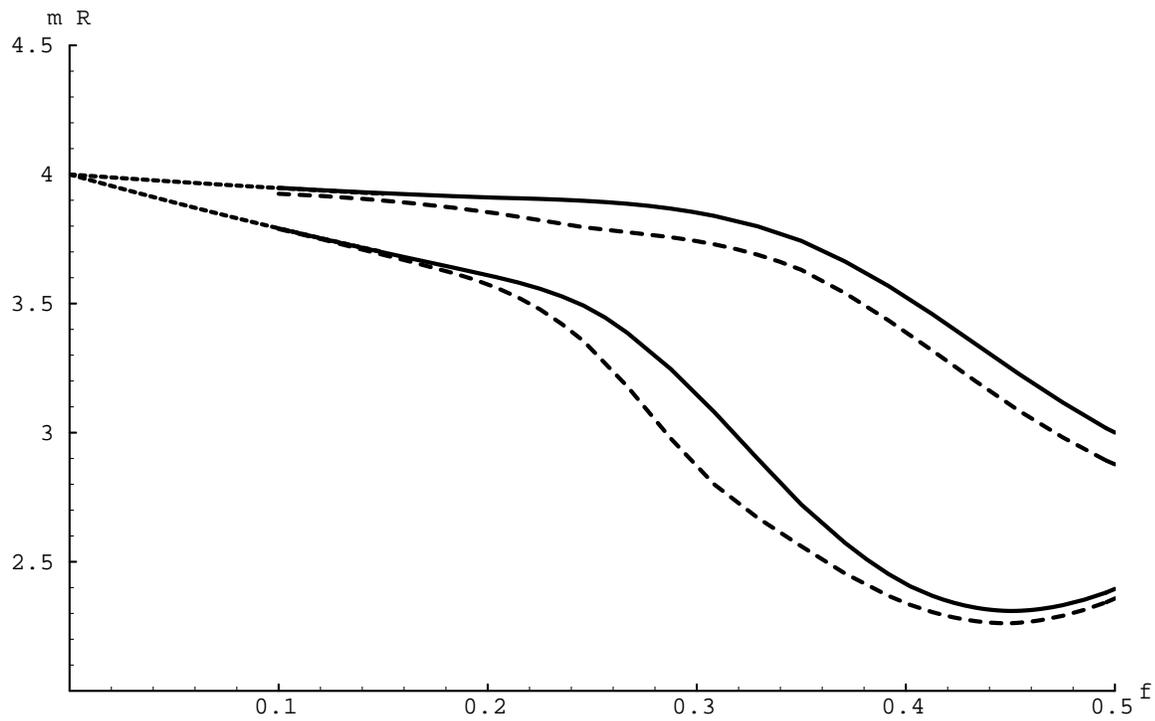}
\caption{Glueball masses for $\theta = 0$ as a function of
the coupling constant.
The drawn curves are the masses of resp. the first scalar ($0^+$) and
tensor ($2^+$) glueball. The dashed lines denote the lower bounds, the dotted
lines the perturbative result.}
\label{massas}
\end{figure}
\begin{figure}[t]
\epsfxsize=152mm
\epsffile{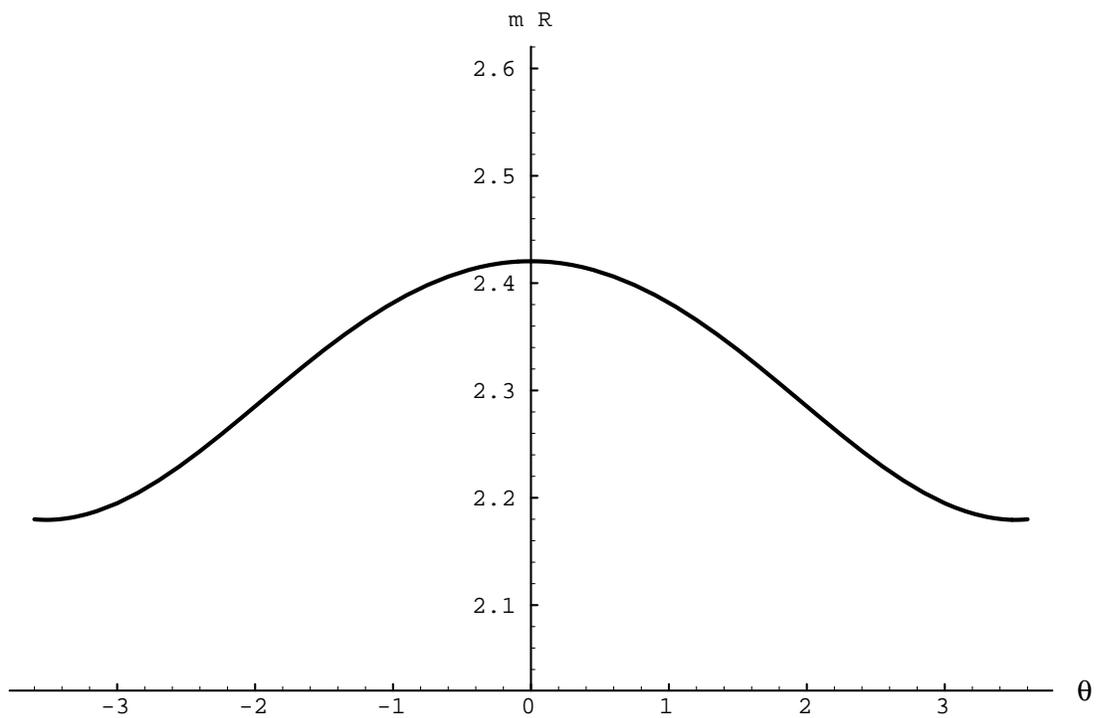}
\caption{Scalar glueball mass at $f=0.4$ as a function of $\theta$.}
\label{theta}
\end{figure}

We found that for the window of coupling constants for which
we trust our approximation, the lowest-lying scalar ($j = 0$)
and tensor ($j=2$) levels are found in respectively
the sectors $(0,0,0)$ and $(1,1,2)$.
See figures~\ref{niveaux} and~\ref{massas}.
For $\theta = 0$, the vacuum corresponds to the ground state of
the $(0,0,0)$-even sector.
The scalar glueball $0^+$ can be identified with the
first excited state in the $(0,0,0)$-even sector.
Note that for small $f$ the lowest
level in the $(0,0,0)$-odd sector $0^-$ has actually a lower energy than
the scalar glueball.
For small $f$, there is virtually no dependence of the masses on $\theta$,
whereas for larger $f$ the $\theta$-dependence is shown in
figure~\ref{theta}.
\section{Conclusions}
Using the Rayleigh-Ritz method we can determine the spectrum of
the effective hamiltonian. The use of Temple's inequality gives
us confidence that our results are accurate,
especially since experience tells us that the actual error is usually
much smaller. The results are also consistent with an ordinary
perturbative calculation of the levels that was done up to
$\Order{f^3}$.
When we make a preliminary comparison with results on the torus,
we find that the mass ratio $m_{2^+}/m_{0^+}$ comes out
rather nice: for a volume of $1~\fm^3$,
corresponding roughly to $f = 0.4$, we obtain the value $1.5$.
\par
Our model is valid for values of the coupling constant at which
only the boundary conditions at and near the sphalerons are felt.
Checking this a posteriori with the help of plots of the wave function in the
$(r_c,r_d)$ plane indicates that $f$ should not be larger than roughly $0.4$.
Consider the function
\beq
  |\psi(r_c,r_d)|^2 =  \int d \ch~d \dh~|\psi(c,d)|^2
\eeq
which is a measure of the
probability distribution in the $(r_c,r_d)$ plane.
Dividing this $|\psi(r_c,r_d)|$ by $r_c^4 r_d^4$, we obtain
a function with the expected behaviour of the true wave function:
it is localised at the sphalerons and decays exponentially in
the transverse directions.
Although the lowest barrier is the sphaleron,
the volume factor $r_c^4 r_d^4$ causes the configurations that
are close to the sphaleron but have a somewhat higher energy
to make the dominant contribution to the tunnelling.
The relevant parameter here is the characteristic decay length
of the wave function, which in turn is determined by the rise of the
potential in the transverse directions.
\par
Although we are primarily interested in the case $\theta = 0$,
an appreciable dependence on $\theta$ for a certain value of $f$
shows that the non-perturbative influence of the boundary has
become important.
To explain the fact that the spectrum is not exactly periodic
in $\theta$, note that our implementation of the
$\theta$-dependence~(\ref{alphadef}) only has this periodicity in
the limit $\beta \rightarrow \infty$. The volume effect
described above implies that the
relevant distribution $|\psi(r_c,r_d)|$ samples a piece of
the boundary over which the phase
difference already starts to depart from $e^{i \theta}$.
\par
The first improvement that has to be made is to take the
higher order corrections of the other modes into account.
The leading order taken here gives rise to a renormalisation of
the coupling constant. Higher order corrections may alter the
potential in our effective model and can for instance result
in a different degree of localisation around the sphalerons.
We hope to report on these matters in the near future.

\section{Acknowledgment}
The author would like to thank Pierre van Baal for many
helpful discussions on the subject.


\newpage

\end{document}